\documentclass[%
reprint,
superscriptaddress,
nofootinbib,
amsmath,amssymb,
aps,longbibliography,
]{revtex4-2}

\usepackage{graphicx}%
\usepackage{verbatim}
\usepackage[version=4]{mhchem}
\usepackage{xcolor}
\usepackage{physics}
\usepackage[normalem]{ulem}
\usepackage{hyperref}
\usepackage{pdfpages}
\usepackage{pgffor}
\usepackage{booktabs}

\usepackage[capitalise]{cleveref}
\usepackage{bm}%

\usepackage{tabularx}

\makeatletter
\AtBeginDocument{\let\LS@rot\@undefined}
\makeatother

\newcommand{\LiPS}{\ce{Li3PS4}}
\newcommand{\lips}{\LiPS}

\usepackage[normalem]{ulem}
\definecolor{tangerine}{rgb}{0.944,0.522,0}
\definecolor{verde}{rgb}{0.,0.6,0}
\definecolor{rosso}{rgb}{0.9,0.0,0.2}
\definecolor{magenta}{rgb}{0.9,0.2,0.9}

\newif\ifhighlight

\newcommand{\highlight}{\highlighttrue}
\highlight
\newcommand{\editor}[2]{%
  \expandafter\newcommand\csname #1note\endcsname[1]{%
    \textcolor{#2}{(\textbf{#1note:} \textsc{##1})}}%
  \expandafter\newcommand\csname #1\endcsname[1]{%
    \ifhighlight\textcolor{#2}{##1} \else ##1\fi}%
  \expandafter\newcommand\csname #1cancel\endcsname[1]{%
    \ifhighlight\textcolor{#2}{\sout{##1}}\fi}%
  \expandafter\newcommand\csname #1change\endcsname[2]{%
    \ifhighlight\textcolor{#2}{\sout{##1} ##2}\else ##2\fi}%
  \newenvironment{#1text}{\ifhighlight\color{#2}\fi}{\color{black}}
}

\editor{FG}{red}
\editor{resub}{blue}
\editor{DT}{verde}

\begin{document}

\title{
Mechanism of charge transport in lithium thiophosphate
}

\makeatletter
\renewcommand*{\@fnsymbol}[1]{\ensuremath{\ifcase#1\or \dagger\or * \or \ddagger\or
   \mathsection\or \mathparagraph\or \|\or **\or \dagger\dagger
   \or \ddagger\ddagger \else\@ctrerr\fi}}
\makeatother
   
\author{Lorenzo Gigli}
\thanks{These authors equally contributed to this work}
\affiliation{Laboratory of Computational Science and Modeling, Institut des Mat\'eriaux, \'Ecole Polytechnique F\'ed\'erale de Lausanne, 1015 Lausanne, Switzerland}

\author{Davide Tisi}
\thanks{These authors equally contributed to this work}
\affiliation{Laboratory of Computational Science and Modeling, Institut des Mat\'eriaux, \'Ecole Polytechnique F\'ed\'erale de Lausanne, 1015 Lausanne, Switzerland}

\author{Federico Grasselli}
\affiliation{Laboratory of Computational Science and Modeling, Institut des Mat\'eriaux, \'Ecole Polytechnique F\'ed\'erale de Lausanne, 1015 Lausanne, Switzerland}

\author{Michele Ceriotti}
\email{michele.ceriotti@epfl.ch}
\affiliation{Laboratory of Computational Science and Modeling, Institut des Mat\'eriaux, \'Ecole Polytechnique F\'ed\'erale de Lausanne, 1015 Lausanne, Switzerland}

\date{\today}%

\begin{abstract}
{
\textbf{\small{ABSTRACT}}: Lithium ortho-thiophosphate (\LiPS) has emerged as a promising candidate for solid-state-electrolyte batteries, thanks to its highly conductive phases, cheap components, and large electrochemical stability range. Nonetheless, the microscopic mechanisms of Li-ion transport in \LiPS\,are far to be fully understood, the role of \ce{PS4} dynamics in charge transport still being controversial.
In this work, we build machine learning potentials targeting state-of-the-art DFT references (PBEsol, r$^2$SCAN, and PBE0) to tackle this problem in all known phases of \LiPS~ ($\alpha$, $\beta$ and $\gamma$), for large system sizes and timescales.
We discuss the physical origin of the observed superionic behavior of \LiPS: the activation of \ce{PS4} flipping drives a structural transition to a highly conductive phase, characterized by an increase of Li-site availability and by a drastic reduction in the activation energy of Li-ion diffusion. We also rule out any paddle-wheel effects of \ce{PS4} tetrahedra in the superionic phases--previously claimed to enhance Li-ion diffusion--due to the orders-of-magnitude difference between the rate of \ce{PS4} flips and Li-ion hops at all temperatures below melting.
We finally elucidate the role of inter-ionic dynamical correlations in charge transport, by highlighting the failure of the Nernst-Einstein approximation to estimate the electrical conductivity. 
Our results show a strong dependence on the target DFT reference, with PBE0 yielding the best quantitative agreement with experimental measurements not only for the electronic band-gap but also for the electrical conductivity of $\beta$- and $\alpha$-\LiPS.
}
\end{abstract}

\maketitle

\section{Introduction}

The growing demand for portable electronic products and electric vehicles has stimulated the creation of energy storage systems that offers better safety and higher energy density than current Li-ion battery systems\cite{Kwade2018}. 
While commercial Li-ion batteries use organic liquid electrolytes and additives to achieve a high working voltage \cite{AURBACH2000206, Etacheri2011}, these materials pose safety concerns due to their flammability and susceptibility to thermal runaway \cite{ARBIZZANI20114801,Murmann_2015}. 
To address these issues, researchers are developing all-solid-state batteries (ASSBs) with inorganic solid electrolytes (SEs) to provide a sustainable solution for energy storage, exploiting their expected longer lifespan and improved energy efficiency \cite{Janek2016,AHNIYAZ2021100070}. 
Many families of SEs have been considered and studied during these years \cite{Bachman2016inorganic,Famprikis2019,castelli_role_2020}. Sulfides are recognised as uniquely promising materials due to their remarkable mechanical stability and room-temperature ionic conductivity \cite{Seino2014,Kato2016,SAKUDA2017108,TAKADA201874,Hakari2017,materzanini_high_2021}.
In particular, the family of lithium thiophosphates (LPS), with its archetypal \ce{Li3PS4} compound, is widely recognised as one of the most promising family of sulfide electrolytes and it has been the subject of many experimental and computational studies \cite{Kwade2018,deshpande2020roadmap,li2021advance,KUDU2022168,HOMMA201153,kimura2023,de2018analysis,forrester2022,ariga2022new,staacke2021role,Staacke2022,guo_artificial_2022}. 

\ce{Li3PS4} has three main polymorphs: $\alpha$-\ce{Li3PS4} (with space group $Cmcm$ \cite{kaup_impact_2020}), $\beta$-\ce{Li3PS4} ($Pnma$) and $\gamma$-\ce{Li3PS4} ($Pmn2_1$). 
Whereas the $\gamma$ polymorph is the most stable at room temperature, it also exhibits low room-temperature ionic conductivity ($\approx 3\times 10^{-7}~$S cm$^{-1}$ \cite{HOMMA201153}). The system transforms into the metastable $\beta$-polymorph at $573~$K and then into the $\alpha$-polymorph at $746~$K \cite{HOMMA201153}. 
Despite their great relevance, in the past years computational studies have been limited by the use of empirical potentials \cite{ariga2022new,forrester2022} and \textit{ab-initio} molecular dynamics (AIMD) \cite{de2018analysis}, based on density functional theory (DFT) with generalised gradient approximation (GGA) \cite{perdew1996generalized,perdew1996generalizedHole}. 
The former can provide useful mechanistic insights, but fail to correctly predict the activation energies of the conductive phases \cite{forrester2022} and are inherently limited in their accuracy and transferability. Quantum mechanical approaches, on the other hand, are more accurate but they are burdened by a higher computational cost, which hinders their applicability to realistic systems. For example, recent studies based on AIMD-PBE simulations attributed the superionic conductivity of glassy \ce{75Li2S}–\ce{25P2S5} and that of bulk $\beta$-\LiPS~to the presence of fast cation-polyanion correlations -- the so-called paddlewheel effect \cite{zhang_targeting_2020, smith_low-temperature_2020}. While providing evidence of this effect, the simulations carried out in these works are clearly limited in the simulation times and system sizes they can achieve, potentially leading to unphysical outcomes.

\begin{figure*}
    \centering
    \hspace*{-0.7cm}
    \includegraphics[width=1.0\textwidth]{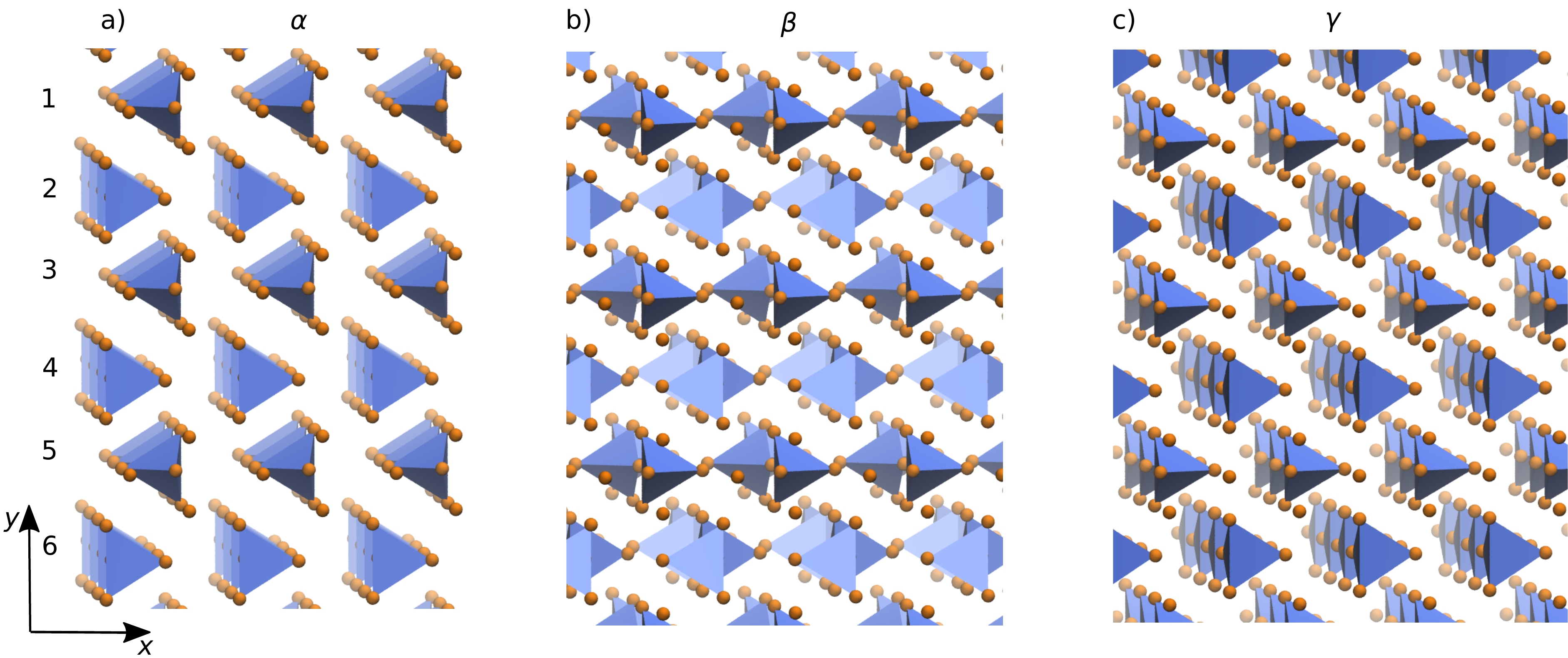}
    \caption{Sketch of the $\alpha$, $\beta$ and $\gamma$ phases of \LiPS, showing the difference of the relative alignment of the \ce{PS4} tetrahedra along reference (010) crystallographic planes (here numbered between 1 and 6 for clarity). The $\gamma$ structure has all tetrahedra aligned along the [100] direction for all this set of planes, while the $\beta$ structure has the tetrahedra aligned along both the [100] direction and the [$\bar{1}00$] across each plane. Finally, the $\alpha$ phase has a staggered ordering: the tetrahedra are aligned along [100]/[$\bar{1}00$] for the planes that are numbered with even/odd numbers.}
    \label{fig:lips-phases}
\end{figure*} 

In the last decade, the advent of machine learning has allowed the construction of interatomic potentials possessing quantum mechanical accuracy at a cost that is only marginally higher than that of classical force fields \cite{Deringer2021,WANG2018178,zeng2023deepmdkit,Bartok2010, Behler2007, Smith2017, schutt2022schnetpack, Rupp2012,Butler2018,fourGenBeheler,Unke2019, Batzner_NatCommun_2022_v13_p2453,PhysRevB.104.104309}. Machine learning potentials (MLP) rely on the construction of physically-motivated representations to predict a given target property. 
In particular, representations of atomic configurations should preserve key physical symmetries: global translational and rotational invariance, as well as invariance with respect to the permutation of atoms of the same chemical species \cite{Willat2019}. 
Among the numerous potential representations, the Smooth Overlap of Atomic Positions (SOAP) \cite{Bartok2013} used in combination with appropriate regression schemes has facilitated the development of ML potentials for simulating a variety of materials properties via extensive finite-temperature thermodynamic sampling \cite{ImbalzanoGaAs2021,Natasha2021,Gigli2022, gigli_modeling_2023,staacke2021role,Staacke2022,Deringer2017,Bartok2018,maresca2018screw,Sivaraman_2020,Deringer2020,DeringerCaroCsanyi,Rosenbrock2021} 
Notable examples of the use of MLPs to study the ionic conductivity in solid-state electrolytes are the Gaussian Approximation Potential (GAP) for lithium thiophosphate developed by Staacke \textit{et al.} \cite{Staacke2022} and the Deep Neural Network (DNN) for \ce{Li10GeP2S12}-type compounds developed by Huang and coworkers \cite{Huang2021}.
Both these studies were able to characterize the diffusion properties of their respective target compounds, and overcome some known limitations of AIMD, namely the small size and the short simulation times that are accessible by this type of modelling. 
Despite these important breakthroughs, two main aspects are still missing to provide a comprehensive study of transport properties in this class of materials. 
First, the accuracy of the aforementioned potentials is limited to the GGA level of theory, due to the choice of the reference DFT functional (PBE and PBEsol) for the calculation of the training set structures. 
While this is a standard choice for performing first-principles calculations in solids, the relatively small number of reference single-point calculations (usually a few thousands) that are needed to reach the desired target ML accuracy enables the use of more accurate references, like meta-GGA and hybrid functionals \cite{perdew_jacobs_2001, becke_new_1993}. 
To our knowledge, no systematic study comparing different DFT references exists up to date for this class of materials. 
Secondly, these studies neglect the contribution of inter-ionic correlations to the electrical conductivity, and its relation with polyanion rotations.

In this work, we train three MLPs to investigate the physical mechanisms of charge transport in \LiPS~and their effect on the electrical conductivity in its stable polymorphs. Each potential is trained over datasets computed at a different level of theory: GGA, metaGGA and hybrid functionals. 
In particular, we use the Perdew-Burke-Ernzerhof functional revisited for solids (PBEsol) \cite{perdew1996generalized,PerdewPBEsol}, the regularized version of the strongly constrained and appropriately normed (r$^2$SCAN) functional \cite{r2SCAN} and the PBE0 functional \cite{adamo1999toward}. 
We explore the temperature dependence of the {ionic} conductivity of \LiPS~showing that different functionals predict different critical temperatures for the onset of the conductive regime, which is roughly associated with the onset of a structural phase transition. 
We also elucidate the importance of including the effects of the interionic correlation in the conductivity, by computing it with the full Green-Kubo (GK) theory of linear response \cite{Green,Kubo}, instead of the Nernst-Einstein approximation commonly employed in the literature. 
Overall, we find that the PBE0 functional gives the best quantitative agreement with existing experimental measurements of the {ionic} conductivity of $\beta$-\LiPS. 
Furthermore, we relate the onset of the superionic phase of the \LiPS~compound with the $\textrm{PS}_4$ flipping dynamics and find that discrete P-S flips induce a structural phase transition from the non-conductive $\gamma$ to a mixture of the $\beta$ and $\alpha$ structures, that cannot be fully resolved at the size and time scale of these simulations. 
This structural change determines a drastic decrease of the slope of the Arrhenius curve and thus a significant reduction of the activation energy of Li-ion diffusion (by a factor of six compared to the $\gamma$ phase). 
Finally, we detect a second transition to a disordered phase with freely rotating polyanions at even higher temperatures that we attribute to the melting of the $\textrm{PS}_4^{3-}$ sublattice. 
Both the transition to the conductive phase and the \LiPS~melting appear as peaks of the heat capacity and are thus associated with separate first-order phase transitions of this material.

\section{Methods}
\label{sec:Methods}

\subsection{Training set construction and validation of the ML models}
\label{sec:MLPs}

\begin{figure}
    \centering
    \hspace{-1.2cm}
    \includegraphics[width=1.1\linewidth]{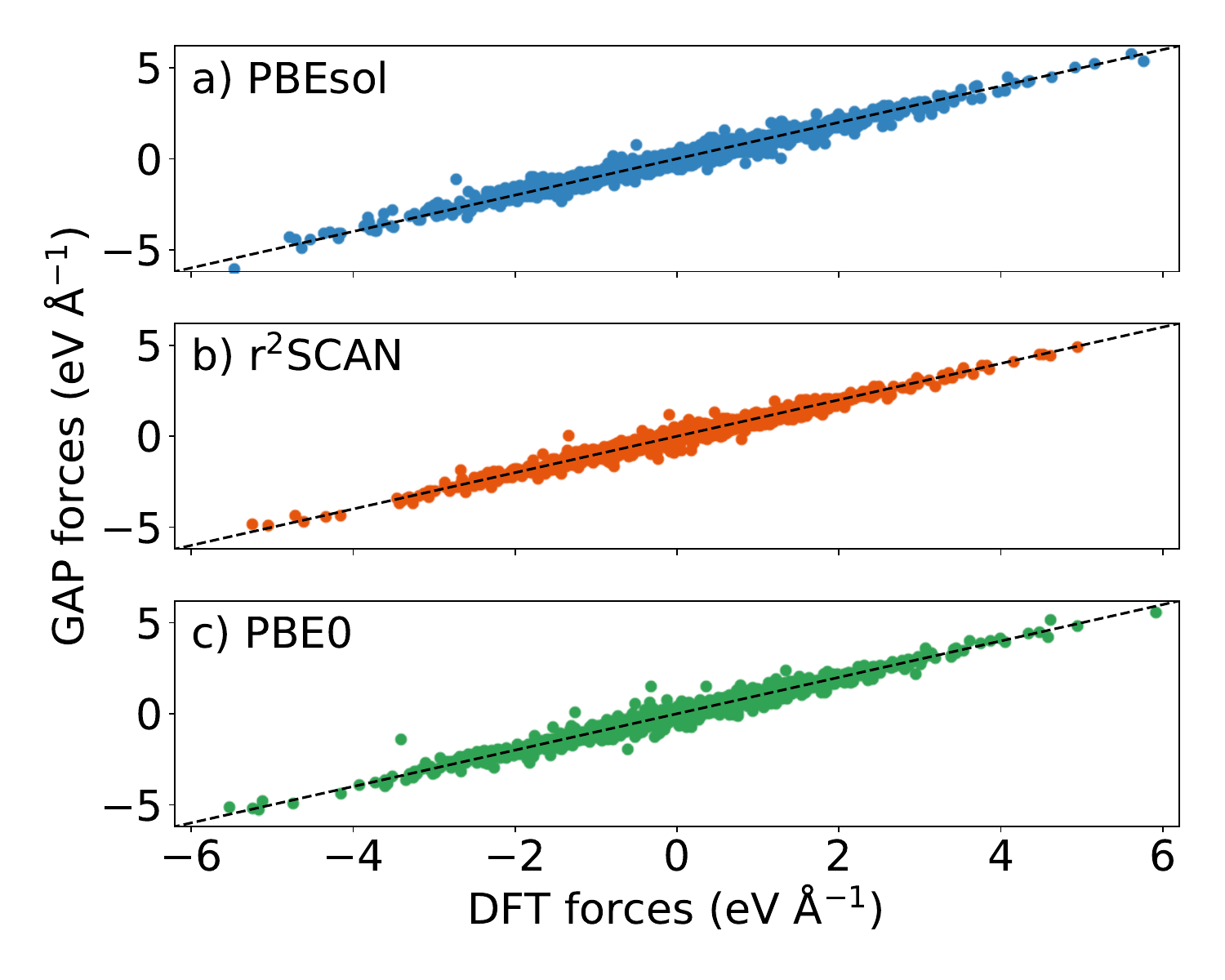}
    \caption{Parity plot of the atomic forces for each model: a) PBEsol; b) r$^2$SCAN; c) PBE0. }
    \label{fig:GAPvsDFT}
\end{figure}

We construct the training set for the ML models in an iterative fashion. A starting dataset of structures is generated by running NVT Car-Parrinello Molecular Dynamics \cite{car_unified_1985} with the PBEsol functional \cite{PerdewPBEsol} as provided by the \textsc{Quantum ESPRESSO} package \cite{gian+09jpcm, giannozzi_advanced_2017,Urru2020,CarmineoQE}, for a set of selected temperatures ($250\,$K, $500\,$K and $1000\,$K) and volumes. This initial dataset is then computed via more converged single-point DFT calculations with the PBEsol functional. Further details can be found in the Supporting Information.

As a next step, we fit a preliminary MLP on this dataset and run finite-temperature Molecular Dynamics (MD) with i-PI \cite{kapi+19cpc} over the entire temperature range of interest (between 200 and 1000 K). 
Among the uncorrelated structures generated in the resulting trajectories, only a subset consisting of the most diverse ones according to the Farthest-Point Sampling (FPS) method \cite{imba+18jcp} is selected and recomputed using DFT.
This active-learning loop, consisting of the regression of an MLP, MD simulations and re-calculation of a set of structures by DFT, allows us to extend the dataset until the model is deemed sufficiently accurate and robust.  
In order to generate datasets for the ML-SCAN and ML-PBE0 models, we select via FPS a subset of snapshots out of the whole PBEsol dataset and we use a two-level machine learning (2LML) scheme \cite{Zaspel2019} to train accurate potentials from a minimal number of expensive r$^2$SCAN or PBE0 calculations. The 2LML is a specific case of the general multilevel machine learning scheme, and consists in: training a ML model on the large PBEsol dataset, then computing energy and forces at the r$^2$SCAN (PBE0) level and finally training a new ML potential on the difference between the ML-PBEsol predicted energies and forces and the true r$^2$SCAN (PBE0) references \cite{Lilienfeld2015,Zaspel2019}.
The final datasets consists of $2400$ structures for the ML-PBEsol model, $740$ for the ML-$\text{r}^2\text{SCAN}$ model and $790$ structures ML-PBE0 model. 
Within these datasets, a subset of $100$ randomly selected structures is used as a test set for the ML-PBEsol model, while a subset of $40$ structures is used for both the ML-$\text{r}^2\text{SCAN}$ and the ML-PBEsol model. PBEsol calculations are performed with \textsc{Quantum ESPRESSO}, while r$^2$SCAN and PBE0 calculations are performed with VASP \cite{VASP1,VASP2,VASP3}. The training of all the models is performed targeting the cohesive energies to avoid offset issues induced by different pseudopotentials. 
Figure \ref{fig:GAPvsDFT} shows the parity plot for the forces of the three models over their respective test sets.
\cref{tab:RMSE} contains the root-mean-square-errors (RMSE) for all models, showing that our model can achieve errors similar (or better) than those obtained in other similar works \cite{staacke2021role,Staacke2022,Huang2021}. The learning curves for each of these models are reported in the Supporting Information {(Sec.~I)}. 
{The Supporting Information also reports results from kernel principal component analysis \cite{scholkopf1997kernel} (Sec.~II) and the newly introduced \textit{local prediction rigidity} \cite{chong2023robustness} (Sec.~III) to check the distribution of the environments in our dataset and along the MD trajectories, and to verify that our training set can reliably represent the complex local environments that occur during PS$_4$ flips. Dynamical properties, like the mean square displacement and atomic diffusivity of Li ions, also appear to be properly reproduced by our ML potentials, as we have directly tested via MD simulations of the $\alpha$ phase at high temperature, obtained with PBEsol \textit{ab initio} potential and with its corresponding ML model, as reported in Sec.~IV of the Supporting Information.}

\begin{table}[h!]
    \centering
    \begin{tabularx}{0.7\columnwidth}{ccc}
    \toprule
       model & RMSE & RMSE (\%RMSE) \\
       & [meV/atom] & [meV/\AA]\\
       \midrule 
       PBEsol  & 7.0 & 120 (14.8\%) \\
       r$^2$SCAN & 6.5 & 141 (15.8\%) \\
       PBE0 & 6.5 & 165 (17.2\%)\\
       \bottomrule 
    \end{tabularx}
    \caption{Root-mean-square-error (RMSE) of the energy (second column) and forces (third column) for all the models. In the third column, the number in parentheses indicates the \%RMSE relative to the standard deviation of forces within the test set.}
    \label{tab:RMSE}
\end{table}

\subsection{Collective variables for \ce{Li3PS4}}
\label{sec:CVs}

The three main polymorphs of \LiPS~are differentiated by the relative orientation of the \ce{PS4} tetrahedra. In order to distinguish them and identify phase transitions in MD simulations, we construct two collective variables (CVs), based on the alignment, along the [100] direction, of the tetrahedra of the (010) planes (see Fig.~\ref{fig:lips-phases}). To this aim, we first compute the polar angle $\theta_\mathrm{SP}$, spanned by the vector $\mathbf{r}_\mathrm{SP} \equiv \mathbf{r}_\mathrm{S} - \mathbf{r}_\mathrm{P}$ that connects, for any \ce{PS4} group, a given S atom with the central P atom, with respect to the x-axis shown in Fig. \ref{fig:PS-flip}:

\begin{equation}
    \label{eq:theta_SP}
     \theta_\mathrm{SP} \equiv \frac{180^{\circ}}{\pi} \arccos \left(\frac{\mathbf{r}_\mathrm{SP} \cdot \hat{\mathbf{x}}}{|\mathbf{r}_\mathrm{SP}|}\right)
\end{equation}

One can then define an order parameter that measures the average alignment of \ce{PS4} tetrahedra within each (010) plane, as follows:

\begin{equation}
    s^C = \frac{1}{4N_\mathrm{P}^C} \sum_{p \in C} \sum_{\langle s,p \rangle} \cos^5 \left(\theta_{sp}\right)
\label{eq:CVC}
\end{equation}
where $C=1,\ldots,6$ labels the (010) planes as shown in Fig. \ref{fig:lips-phases}, $N_\mathrm{P}^C = 16$ is the number of P atoms in each plane, and $\langle s,p \rangle$ represents the S atoms that are first nearest neighbours of P atom $p$. In other words, the outer sum runs over P atoms that belong to a given crystallographic plane, while the inner sum runs on the four S atoms that belong to the tetrahedron centered around atom $p$. Whenever one tetrahedron is perfectly aligned along $\pm x$, the cosine of one P-S angle is close to $\pm 1$, while the remaining three have a value of $\cos(109.5^{\circ}) \approx -0.3338$ or $\cos(70.5^{\circ}) \approx +0.3338$ for $+x$ and $-x$ respectively, due to tetrahedral symmetry. We thus raise the cosine to the fifth power in Eq.~\eqref{eq:CVC}, so that the result of the inner sum will be approximately equal to $\pm 1$. While any odd power greater that $1$ would serve this scope,\footnote{This observation stems from the symmetry of the tetrahedra, $\sum_{\langle \mathrm{s,p} \rangle} \cos \left(\theta_{\mathrm{sp}}\right)=0$} the fifth power gives the best results on preliminary tests.

Since the final aim is to define a global measure of the relative alignment across planes, so as to capture the staggered ordering of the $\alpha$ structure, we construct two intermediate order parameters $s_{\mathrm{even}}$ and $s_{\mathrm{odd}}$, by averaging $s^C$ for even and odd values of $C$. 
We finally construct the following pair of collective variables:

\begin{equation}
    s_1 = \frac{1}{4} \sqrt{s_{\mathrm{even}}^2 + s_{\mathrm{odd}}^2} \tanh(50 s_{\mathrm{even}}) \tanh(50 s_{\mathrm{odd}})
\label{eq:CV1}
\end{equation}

\begin{equation}
    s_2 = \frac{1}{N^C} \sum_C s^C
\label{eq:CV2}
\end{equation}

In Sec.~\ref{sec:phase-trans} we demonstrate that $s_1$ completely distinguishes the three polymorphs, as it takes on values close to $+1$ when the structure is similar to the perfect $\gamma$, $-1$ when it is similar to $\alpha$ and 0 when the structure is close to $\beta$ or fully disordered. $s_2$ instead measures a global alignment of the entire structure and does not resolve the $\beta$ and $\alpha$ structure, as they both contain mixed \ce{PS4} orientations in equal proportions. Still, this CV is meaningful when combined with $s_1$ as it carries information about the relative number of tetrahedra that are aligned along the positive and negative direction of the x-axis. For instance, it can distinguish between two perfect $\gamma$ structures that are mirror-symmetric with respect to a (100) plane.

\subsection{Green-Kubo theory}
\label{sec:GKtheory}

The Green-Kubo (GK) theory of linear response \cite{Green,Kubo} provides a rigorous and elegant framework to compute transport coefficients of extended systems in terms of the stationary time series of suitable fluxes evaluated at thermal equilibrium with MD. 
For an isotropic system of $N$ interacting particles, the GK expression for the electrical conductivity reads:
\begin{equation}\label{eq:GKeq}
    \sigma=\frac{\Omega}{3k_BT}\int_0^{\infty} \langle \mathbf{J}_q(\Gamma_t) \cdot \mathbf{J}_q(\Gamma_0) \rangle\, dt,
\end{equation}
where $k_B$ is the Boltzmann constant, $T$ the temperature and $\Gamma_t$ indicates the time evolution of a point in phase space from the initial condition $\Gamma_0$, over which the average $\langle \cdot \rangle$ is performed. $\mathbf{J}_q$ is the charge flux, that can be easily computed from MD, knowing the velocities of the atoms, $\mathbf{v}_i$, and their charges, $q_i$:
\begin{equation}\label{eq:elec-currents}
    \mathbf{J}_q=\frac{e}{\Omega}\sum_{i} q_i \mathbf{v}_i .
\end{equation}
Here, the sum runs over all the atoms, $e$ is the electron charge, and the $q_i$ are equal to the nominal oxidation number of the atoms \cite{Grasselli2019}: {in the absence of electronic conductivity due to conduction electrons or polaronic states, the overall electrical conductivity coincides with that obtained from \cref{eq:elec-currents} using integer, time-independent ionic charges \cite{Pegolo2020oxidation}.}

A commonly used approximation of \cref{eq:GKeq} is the Nernst-Einstein (NE) equation:
\begin{equation}\label{eq:NE}
    \sigma_\mathrm{NE} = \frac{e^2 q_{\text{Li}}^2 N_{\text{Li}} }{\Omega k_B T}D_{\text{Li}}
\end{equation}
where $D_{\text{Li}}$ and $N_{\text{Li}}$ represent the diffusion coefficient and the total number of the lithium atoms respectively. 
\Cref{eq:NE} is widely used in practice to estimate the {ionic} conductivity, due to the high statistical accuracy with which atomic diffusion coefficients can be computed from numerical simulations \cite{all-tild90book}. Nevertheless, its application to solid-state-electrolytes (SSEs) is burdened by systematic errors \cite{Marcolongo2017}: in fact, the large interatomic dynamical correlations, both between carriers (Li$^+$) and the solid matrix (\ce{PS4}$^{3-}$) and among the carriers themselves, which is typical in systems with a high carrier concentration like SSEs, is completely neglected by Eq.~\eqref{eq:NE}.  
The discrepancy between $\sigma$ and $\sigma_\mathrm{NE}$ can be quantified by the Haven ratio \cite{MURCH1982177,C3EE41728J,Marcolongo2017}:
\begin{equation}
\label{eq:HR}
    H_R = \frac{\sigma_\mathrm{NE}}{\sigma}
\end{equation}
We redirect the reader to Sec. \ref{sec:elec-conductivity} for a thorough comparison between $\sigma$ and $\sigma_\mathrm{NE}$ in the different polymorphs of \LiPS.

From a methodological standpoint, \cref{eq:GKeq} can be expressed in an equivalent formulation, called the Helfand-Einstein (HE) formula, which exhibits better statistical behaviour \cite{Grasselli2021}: 
\begin{equation}\label{eq:sigmaHE}
    \sigma_\mathrm{HE} = \frac{\Omega}{3k_BT}\lim_{\mathcal{T}\to +\infty} \int_{0}^{\mathcal{T}}\langle \mathbf{J}_q(\Gamma_t) \cdot \mathbf{J}_q(\Gamma_0) \rangle \left( 1-\frac{t}{\mathcal{T}} \right)dt
\end{equation}

The Li diffusivity appearing in Eq.~\eqref{eq:NE} is obtained from the asymptotic slope of the mean square displacement of the Li ions:
\begin{equation}
    D_{\text{Li}} = \frac{1}{6} \lim_{t\to\infty} \frac{d}{dt} \frac{1}{N_\mathrm{Li}}\sum_{i=1}^{N_\mathrm{Li}} \langle |\mathbf{r}_i(t) - \mathbf{r}_i(0)|^2 \rangle
\end{equation}
In this case, care has to be taken to compute $D_{\text{Li}}$ in the reference frame where the solid matrix is fixed to avoid non-physical contribution to the calculations of the electrical conductivity. These spurious effects arise for simulations run in the barycenter reference frame, where the position of the center of mass of the entire system is fixed. In practice, the difference between $D_{\text{Li}}$ computed in these two reference frames vanishes when the box size is increased \cite{Grasselli2022}.

Due to its very general formulation, the GK expression of \cref{eq:GKeq} can be used to investigate, with minimal variations, other characteristic properties of \LiPS. In Sec.~\ref{sec:PS4-rotations} we will characterize the rotational properties of the \ce{PS4} polyanions at high temperature, by computing a rotational diffusion coefficient as follows: 

\begin{equation}
\label{eq:Domega}
    D_\omega = \frac{1}{3N_{\ce{PS}}} \sum_{j=1}^{N_{\ce{PS}}} \int_0^\infty \langle \bm{\omega}_j(t) \cdot \bm{\omega}_j(0) \rangle \, dt\,.
\end{equation}

In this equation $j$ runs over every P-S bond of each \ce{PS4} tetrahedron and $\omega_j (t)$ represents the time series of its angular velocity: 

\begin{equation}
    \bm{\omega}_j(t) \equiv \frac{(\mathbf{r}^j_{\mathrm{P}} - \mathbf{r}^j_{\mathrm{S}}) \cross (\mathbf{v}^j_{\mathrm{P}} - \mathbf{v}^j_{\mathrm{S}})}{|\mathbf{r}^j_{\mathrm{P}} - \mathbf{r}^j_{\mathrm{S}}|^2}
\end{equation}

where $\mathbf{r}^j_{\mathrm{P,S}} (t)$ and $\mathbf{v}^j_{\mathrm{P,S}} (t)$ are the positions and the velocities of the P and the S atom belonging the $j$-th bond. 

\subsection{ML-MD computational details}
We use the MLPs constructed in Sec.~\ref{sec:MLPs} to investigate the physics of \LiPS~via constant-temperature MD simulations using a combination of i-PI \cite{CERIOTTI2014IPI,kapi+19cpc}, \textsc{lammps} \cite{plim95jcp,LAMMPS} and \textsc{librascal} \cite{librascal}. 
In order to simulate the phase transitions and the charge transport in \LiPS, we perform MD simulations in the NpT ensemble of a quasi-cubic 768-atom cell in all stable $\alpha$, $\beta$ and $\gamma$ phases with a constant isotropic pressure of $p = 0$~atm for a set of temperatures between $200$~K and $1000$~K. 
The system's center of mass is kept fixed during the simulations. A generalized Langevin equation (GLE) thermostat \cite{GLEprl,Ceriotti2010} is used to equilibrate the cell volume, while a stochastic velocity rescaling (SVR) thermostat \cite{Bussi2007} is used to thermalize the velocity distribution of the atoms without affecting significantly the dynamical properties. 
The characteristic times of the barostat, the SVR thermostat and the MD timestep are set to $1$~ps, $10$~fs and $2$~fs, respectively. We run these simulations long enough to ensure statistical convergence of the {ionic} conductivity (see Sec.~\ref{sec:GKtheory}). 
Specifically, we run the weakly conductive simulations of the $\gamma$ phase for $\sim 6$~ns, the $\beta$ phase for  $\sim 4$~ns the $\alpha$ phase for $\sim 2$~ns.
As discussed in Sec.~\ref{sec:phase-trans}, this setup ensures that the system can sample configurations within the range of stability of each phase, without explicitly requiring a quantitative prediction of the temperature-dependent phase diagram of \LiPS. A validation of the setup via a heat-quench simulation in the NST ensemble is found in the Supporting Information {(Sec.~VII)}.

\section{Results}

In this section, we compare the results obtained with the PBEsol, PBE0 and $\textrm{r}^2$SCAN functional and the corresponding ML models, as described in Sec.~\ref{sec:Methods}. In subsection \ref{sec:elec-band-struct}, we compute the electronic band structure of the $\beta$ polymorph using DFT and show that the band gap predicted by the PBE0 functional is in good agreement with experiments, while PBEsol and $\textrm{r}^2$SCAN  considerably underestimate it. In the following subsections, we investigate the finite-temperature predictions of the ML models. First, in subsection \ref{sec:phase-trans}, we analyse the MD simulations of the \LiPS~polymorphs using the collective variable introduced in Sec. \ref{sec:CVs} and discuss the onset of a phase transition from the $\gamma$ structure to a structure with a mixed $\alpha$ and $\beta$ arrangement by increasing the temperature. 
In subsection \ref{sec:PS4-rotations}, we investigate the rotational dynamics of the $\textrm{PS}_4$ tetrahedra and relate the occurrence of phase transitions with the thermal activation of correlated polyanion flips. Furthermore, we detect the presence of a high-temperature liquid phase characterized by freely rotating polyanions.
In subsection \ref{sec:elec-conductivity}, we compute the {ionic} conductivity from MD NpT simulations using both the NE and the HE expressions introduced in Sec. \ref{sec:GKtheory} and the Haven ratio of \LiPS. 
Finally, in subsection \ref{sec:spatial-corr}, we discuss the role of spatial correlations and their effect on the calculated ionic conductivity.

\subsection{Electronic band structure}
\label{sec:elec-band-struct}

\begin{figure}
 \centering
\includegraphics[width=\columnwidth]{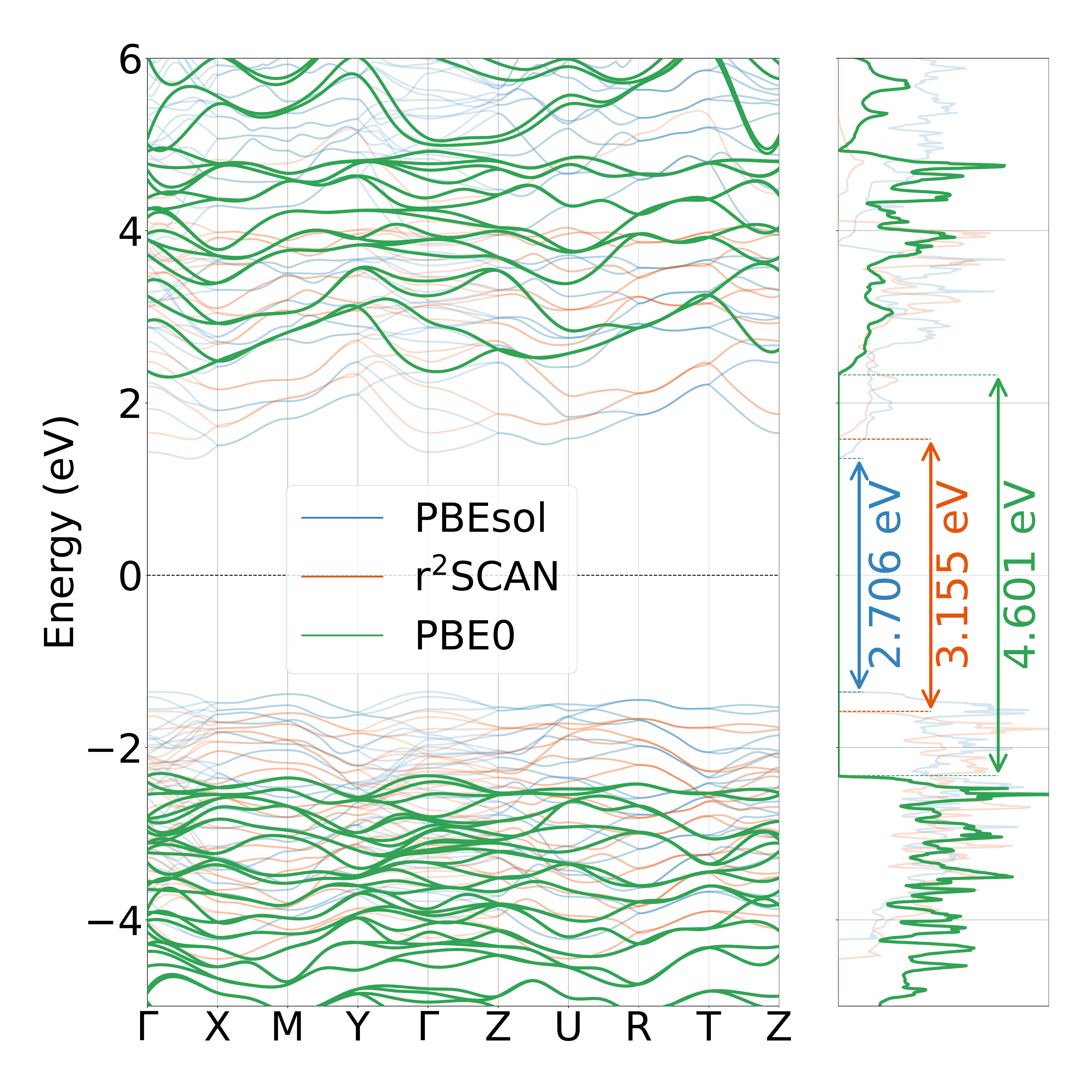}
\caption{Electronic bands and density of states of \ce{Li3PS4} for the $\beta$-phase.}
\label{fig:band_structure}
\end{figure}

\begin{table}
    \centering
    \begin{tabularx}{0.7\columnwidth}{ >{\raggedright\arraybackslash}X 
        >{\centering\arraybackslash}X 
        >{\centering\arraybackslash}X  }
        \toprule
        & $E_g^{\gamma}$ [eV] & $E_g^{\beta}$ [eV] \vspace{0.1cm} \\ 
        \midrule
        PBEsol & 2.649 & 2.706 \\
        $\text{r}^2$SCAN & 3.088 & 3.155 \\
        PBE0 &4.566 & 4.601 \\
        exp. \cite{liu_anomalous_2013,Rangasamy2014} & - & 5 \\
       \bottomrule 
    \end{tabularx}
    \caption{Energy Band Gap with the different models for the $\gamma$ and $\beta$ phase. The PBE0 values are in best agreement with the reported experimental value from Ref.~\cite{Rangasamy2014}.}
    \label{tab:BandGap}
\end{table}

The Generalized Gradient Approximation (GGA) functionals offer a good compromise between accuracy and computational efficiency, making them a practical choice for a broad range of materials and systems.
It is known, however, that they often fall short when it comes to accurately characterizing critical electronic properties, such as the electronic band gap, which is frequently underestimated in GGA, and the density of states \cite{Stroppa_2008,Csonka2009,Hermet2012}.
In order to solve this problem, different functionals have been developed, such as meta-GGA \cite{perdew_jacobs_2001} and hybrid functionals \cite{becke_new_1993}, that offer more accurate approximations of the exchange-correlation functional and are able to better describe long-range electron-electron interactions. These new functionals have enabled in more accurate predictions of electronic properties in a variety of different materials  \cite{Mardirossian2017,yang2023range,Zhang2021,Schmidt2022,Tisi2021,malosso2022} and notably solid-state electrolytes \cite{Marana2022,Lim2018,Pegolo2022}. In \cref{fig:band_structure} we compare the band structure and the density of states (DOS) for the $\beta$ phase computed with the PBEsol (GGA), $\text{r}^2$SCAN (meta-GGA) and PBE0 (hybrid) functionals\footnote{The results for the $\gamma$ phase can be found in the SI}. Since in the $\beta$ phase the Wyckoff sites of the Li atoms have partial occupations, we perform the calculation using one (B3C1 \cite{Lim2018}) of the known configurations with minimum energy, since the electronic bands and the gap are only weakly dependent on this choice \cite{Lim2018}. Furthermore, in \cref{tab:BandGap} we compare the band gaps predicted by the different functionals in the $\gamma$ and $\beta$ phases and recent experimental measurements. We note that both PBEsol and $\text{r}^2$SCAN considerably underestimate the electronic band gap, while PBE0 shows a remarkably good agreement\footnote{We remark that the reported experimental value from Ref.~\onlinecite{Rangasamy2014} is obtained as the electro-chemical window and, as such, represents an upper limit of the band gap.}, thus further motivating the use of this functional as a reference for the training of a dedicated ML model.

\subsection{Structural phase transitions}
\label{sec:phase-trans}

\begin{figure}[h!]
    \centering
    \includegraphics[width=0.5\textwidth]{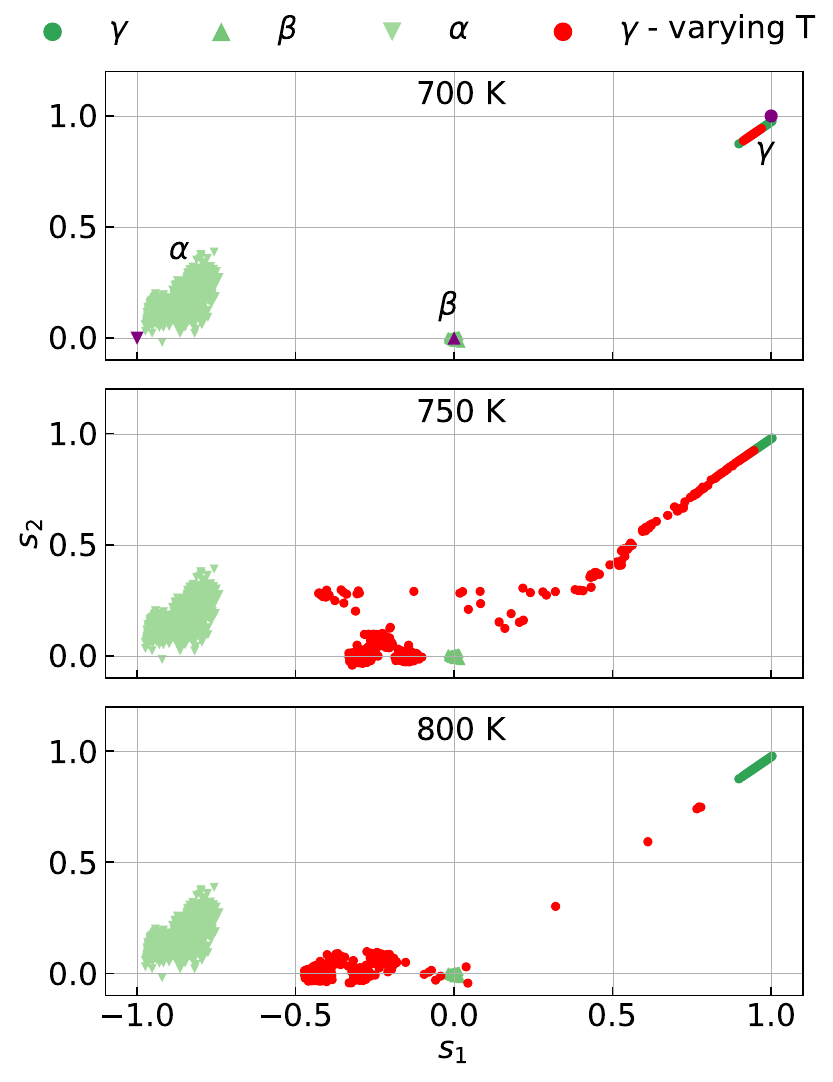}
    \caption{Evolution of the collective variables of $\gamma-$\LiPS (red points) sampled over a set of MD trajectories generated with the ML-PBE0 model. Green markers of three different shades represent a sample of reference points extracted from all MD simulations in the $\alpha$, $\beta$ and $\gamma$ phase below $T_c = 750$\,K where no phase transitions are observed. The purple markers in the topmost panel indicate the CVs for the ideal crystalline structures.}
    \label{fig:gamma-CVs}
\end{figure}

We use the pair of CVs $s_1$ and $s_2$, introduced in Sec.~\ref{sec:CVs}, to investigate the presence of structural phase transitions appearing in the MD trajectories.
As anticipated, $s_1$ characterizes the mutual orientation of adjacent (010) planes (i.e., even and odd numbers in Fig.~\ref{fig:lips-phases}) along the [100] direction. Consequently, $s_1 = -1$ for the $\alpha$ phase, where the \ce{PS4} orientation of adjacent planes is antiparallel, $s_1 = 0$ for the $\beta$ phase (each plane has no net orientation of \ce{PS4} units), and $s_1 = +1$  for the $\gamma$ phase, where adjacent planes share the same orientation of \ce{PS4} tetrahedra along the positive $x$ axis.
Conversely, $s_2$ measures the global orientation of \ce{PS4} tetrahedra and vanishes for both the $\alpha$ and $\beta$ phases, while is equal to 1 for the $\gamma$ phase.
Figure \ref{fig:gamma-CVs} displays, with red dots, the evolution of the CVs across a set of MD simulations run with the ML-PBE0 model at different temperatures $T$, and initialized in the $\gamma$ phase. The green markers of three different shades represent reference points sampled from MD simulations in the $\alpha$, $\beta$ and $\gamma$ phases below $T_c = 750$\,K, and are used as a guide to interpret the $T$-dependent results. For $T < T_c$, the red dots are all concentrated in a region around $(s_1,s_2) = (1,1)$, which is typical of the pure $\gamma$ phase, the small deviations being due to the thermal motion of the atoms only. As $T$ is raised above $\approx~T_c$, a structural transition occurs, and the CVs approach a region in between the $\alpha$ and $\beta$ phases. Although the total timescale and size of the simulations are not sufficient to allow for a complete transition of the $\gamma$ phase to a specific polymorph, but only to intermediate configurations, our MD simulations capture the microscopic driving mechanism, i.e., the onset of concerted reorientations of \ce{PS4} tetrahedra.

\begin{figure}[h!]
    \centering
    \includegraphics[width=\columnwidth]{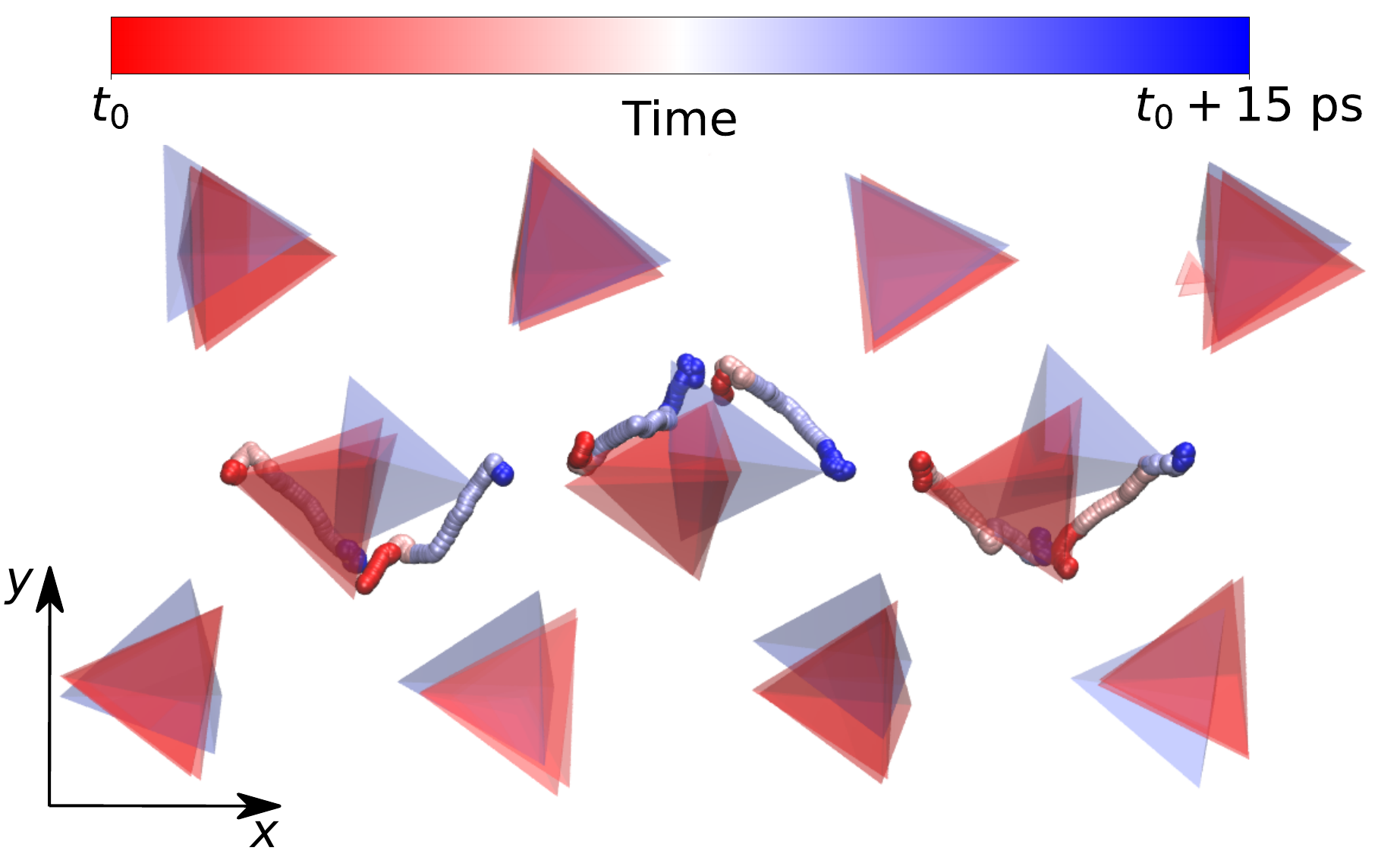}
    \caption{Transition corresponding to the re-orientation of one line of $\textrm{PS}_4$ tetrahedra. Tetrahedra colored in red and blue correspond to snapshots of the solid matrix taken over  the transition for one trajectory at 750 K starting from the $\gamma$ structure and run with the ML-PBE0 model. The trajectory of two vertices of each tetrahedron is displayed with lines, that are colored with an RWB scheme and with a smoothening window of $2.5\,$ps. The red end of these lines corresponds to $t_0$, while the blue end to $t_1=t_0 + 15\text{ ps}$.}
    \label{fig:PS-flip}
\end{figure}

Figure \ref{fig:PS-flip} shows a typical example of this phenomenon, occurring during a $\approx 15$-ps segment of a MD trajectory: the starting (ending) configuration is depicted in red (blue). The color fades from red to blue (with an RWB scheme) in a continuous manner as the transition occurs. The trajectories of the S atoms lying at the vertices of the \ce{PS4} tetrahedra clearly indicate that the reorientation of the entire row occurs coherently, and not as a collection of individual, decorrelated flips. Figure \ref{fig:PS-flip} also shows that the transition is purely orientational, and does not occur through the hopping of S anions between adjacent \ce{PS4} groups. Additional information and a comparison of this mechanism with an heat-quench simulation showing a similar behavior can be found in the Supporting Information {(Secs. VII and VIII)}. 

\subsection{\ce{PS4} rotational dynamics and heat capacity}
\label{sec:PS4-rotations}

\begin{figure*}
    \centering
    \includegraphics[width=0.8\textwidth]{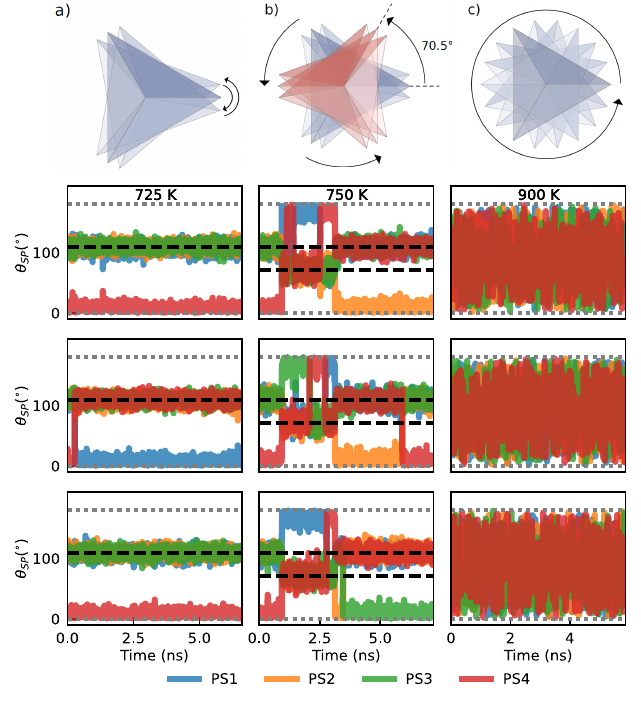}
    \caption{Sketch of the rotational dynamics of the \ce{PS4} groups: low temperature (panel a), where only small librations with respect to the initial configuration occur, at intermediate temperature (panel b) where \ce{PS4} flips determine the structural transition observed in \cref{fig:gamma-CVs} and at high temperature (panel c) where the system is melted. The lower plots show the time evolution of the polar angle $\theta_\mathrm{SP}$ as defined in Eq.~\eqref{eq:theta_SP} (angle with respect to the x-axis) for a set of three distinct tetrahedra forming a [100] row. Each panel represents the dynamics of the four PS bonds forming each tetrahedron. Rows correspond to different tetrahedra, while columns correspond to different NpT trajectories at $T=725$ K, $750$ K and $900$ K. Horizontal dashed black lines indicate the position of the ideal tetrahedral angles at $70.5^{\circ}$ and $109.5^{\circ}$, while grey dotted lines mark the extremes of the domain of $\theta_\mathrm{SP}$ (i.e. $0^{\circ}$ and $180^{\circ}$).}
    \label{fig:Euler-dyn}
\end{figure*}

\begin{figure}
    \centering
    \includegraphics[width=\columnwidth]{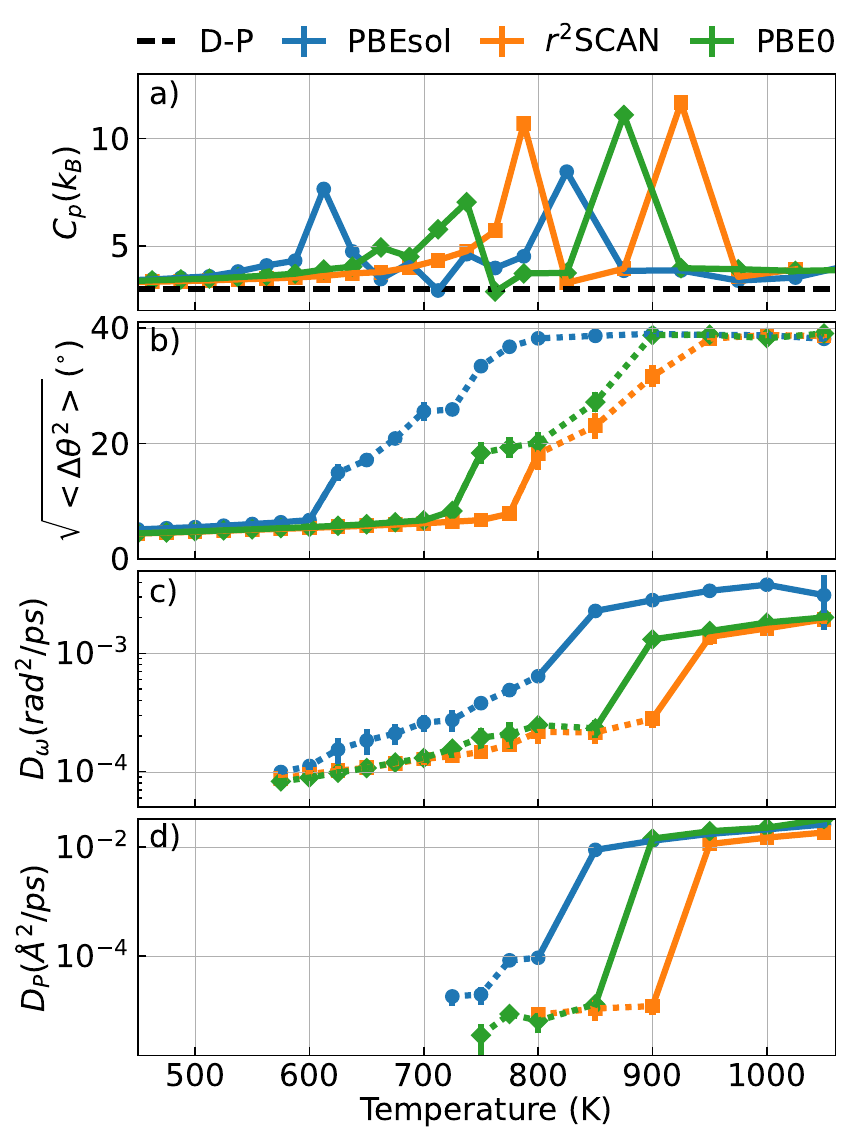}
    \caption{{a)} the heat capacity ($C_p$), {b)} the fluctuations of the P-S polar angle averaged over all bonds of every \ce{PS4} tetrahedron in the simulation box ($\sqrt{\langle \Delta \theta^2 \rangle}$), {c)} the \ce{PS4} rotational diffusion coefficient ($D_{\omega}$) and {d)} the linear diffusion coefficient of the P atoms ($D_\mathrm{P}$) as a function of the simulated temperature. Solid/dashed lines represent each quantity in the temperature regime where it is well/ill-defined. The dashed black line of panel a) represents the Dulong-Petit limit, $3k_B$.}
    \label{fig:heat-cap}
\end{figure}

Further insights into the rotational reorientation of \ce{PS4} planes and their relation with structural transitions can be obtained by a direct investigation of the rotational dynamics of \ce{PS4} tetrahedra, and specifically those that form [100] rows.  \Cref{fig:Euler-dyn} shows the dynamics of the polar angles $\theta_\mathrm{SP}$, as defined in \cref{eq:theta_SP}, for a set of four tetrahedra (row panels) that belong to the same [100] row in NpT simulations run with the ML-PBE0 model\footnote{The ML-PBEsol and the ML-$\text{r}^2$SCAN models give the same qualitative behaviour of the ML-PBE0 model. Notably, all of them display the same phase transitions, albeit at different temperatures (see also Fig.~\ref{fig:heat-cap})}. We also compare three trajectories initialized in the $\gamma$ phase and equilibrated at $T=725$ K, i.e.~just below $T_c=750$ K (left column); at $T=T_c$ (central column); and above melting, at $T=900$ K (right column). The four lines in the plots correspond to the dynamics of the four bonds that constitute each \ce{PS4} tetrahedron.

At 725\,K, only small angular fluctuations occur, with no reorientation of the tetrahedra. The average angles of the P-S bonds define the mean orientation of each tetrahedron: as one of the P-S bonds always has an average value that is close to 0, the orientation is along the positive x-axis ($+x$) during the entire simulation time, which is typical of the $\gamma$ phase. Instead, the angles of the other three bonds oscillate around $\theta_0 = 109.5^{\circ}$ (black dashed line), corresponding to the P-S bond angles in the perfect tetrahedral geometry. The jump observed in the central panel of the left column corresponds to a rotation of the tetrahedron, that does not involve a re-orientation towards the negative x-axis. 

Instead, at the transition temperature ($T_c=750$\,K), simultaneous flips of one P-S bond from 0 to $70.5^{\circ}$ and another P-S bond from $109^{\circ}$ to $180^{\circ}$  correspond to the reorientations of the tetrahedra from $+x$ to $-x$, consistently with the mechanism shown Fig. \ref{fig:PS-flip}. Notably, these flips occur at the same instants of time for every tetrahedron in a row (see, e.g., central column of \cref{fig:Euler-dyn} between 1 and 2.5 ns), confirming that they are highly correlated across [100] rows. This effect is at the basis of the phase transition observed in Fig. \ref{fig:gamma-CVs} as it modifies the relative orientation of tetrahedra across (010) crystallographic planes. It is crucial to note, in this respect, that the spatial correlation of these flips at the transition extends up to the edge of the simulation box, potentially leading to finite-size effects. Specifically, we expect this transition to manifest in larger boxes by nucleation of ordered clusters with opposite orientation, as a result of the formation of defects with sudden changes of the \ce{PS4} orientation. We also note from panels b) of \cref{fig:Euler-dyn} that the time lag between subsequent flips is of the order of 1 ns. As we will see in Sec.~\ref{sec:elec-conductivity}, the presence of this long time scale will be important to elucidate the mechanism of charge transport in this material.

At 900\,K, the dynamics of the tetrahedra changes dramatically and a second phase transition occurs. In particular, all PS bonds across every tetrahedron span the entire range of angles between $0^{\circ}$ and $180^{\circ}$. This suggests that, unlike the phases observed at lower temperatures, the tetrahedra are freely rotating in the simulation box. A direct inspection of the simulations indicates that this behaviour is accompanied by a melting of the system into a mixture of Li$^+$ cations and \ce{PS4}$^{3-}$ anions (see also the P-P and P-S radial distribution functions shown in the {Supporting Information, Fig.~S9}). 

The onset of these two phase transitions--from the non-conductive $\gamma$ phase to the superionic $\beta/\alpha$ hybrid phase and the melting of \LiPS--can be quantitatively investigated for all ML models by computing the temperature dependence of a set of relevant quantities. Panel a) of Fig.~\ref{fig:heat-cap} shows, for each of the ML models, the isobaric heat capacity $C_p(T)$ computed from the finite-difference derivative, with respect to $T$, of the mean enthalpy collected in the NpT simulations. Panel b) shows the temperature dependence of the mean squared fluctuation of the polar angle $\theta_{SP}$ as defined in Eq.~\ref{eq:theta_SP} and further averaged over every P-S bond. Panel c) shows the \ce{PS4} rotational diffusion coefficient $D_\omega$, defined in Eq.~\ref{eq:Domega}, and panel d) the linear diffusion coefficient of the P atoms, $D_\mathrm{P}$.

$C_p(T)$ displays two distinct peaks, characteristic of the phase transitions observed in Fig.~\ref{fig:Euler-dyn}, while the associated critical temperatures depend on the specific functional.
We can characterize more clearly the position of these peaks by analysing their relation with the microscopic quantities shown in panels b), c) and d). Since the first transition is associated with discrete \ce{PS4} flips, the increase of $C_p$ is accompanied by a sudden change of slope of the angular fluctuations. Conversely, the transition to the molten phase occurs with a dramatic increase of both $D_\omega$ and $D_\mathrm{P}$ by one and two orders of magnitude, respectively. In other words, the action of thermal fluctuations at this high temperature destroys the periodic arrangement of P atoms, while the tetrahedra are still intact and can freely rotate by a rate given by $D_{\omega}$. Notably, the transition temperature to the molten salt as predicted by the ML-PBE0 model is in agreement with a previous experimental measurement of the binary phase diagram of $\beta$-\LiPS - \ce{Li4GeS4} solid solutions obtained through differential thermal analysis\cite{hori_phase_2015}. More specifically, the transition point upon heating is reported to be $600^{\circ}\textrm{C}$ when the concentration of $\beta$-\LiPS~is equal to 98$\%$ (P-rich regime), which is compatible with our PBE0 estimate.

We stress that the angular deviations of panel b) can be defined only with respect to a local equilibrium for each P-S bond and are thus meaningful only in the low-T phase. Conversely, the diffusion coefficients of panels c) and d) are physically meaningful when the simulations sample sufficiently many configurations with displaced P atoms and rotated \ce{PS4} anions. They are thus not well defined if the MD simulations are not fully ergodic \cite{Poletayev2022}. In Fig.~\ref{fig:heat-cap}, we display each of the quantities with solid lines in the regions where they are well defined, otherwise they are shown with dotted lines.
The phase transitions investigated so far have strong implications on electrical conduction, as we describe in the following subsection.

\subsection{{Ionic} conductivity and Haven ratio}
\label{sec:elec-conductivity}

The {ionic} conductivity, $\sigma$, is the crucial property to identify promising solid-state electrolytes. As discussed in Sec.~\ref{sec:GKtheory}, the GK theory of linear response, in its HE formulation, gives us an efficient and statistically robust method to obtain an estimate of $\sigma$ from equilibrium MD simulations at any target temperature.

\begin{figure*}
    \centering
    \includegraphics[width=\textwidth]{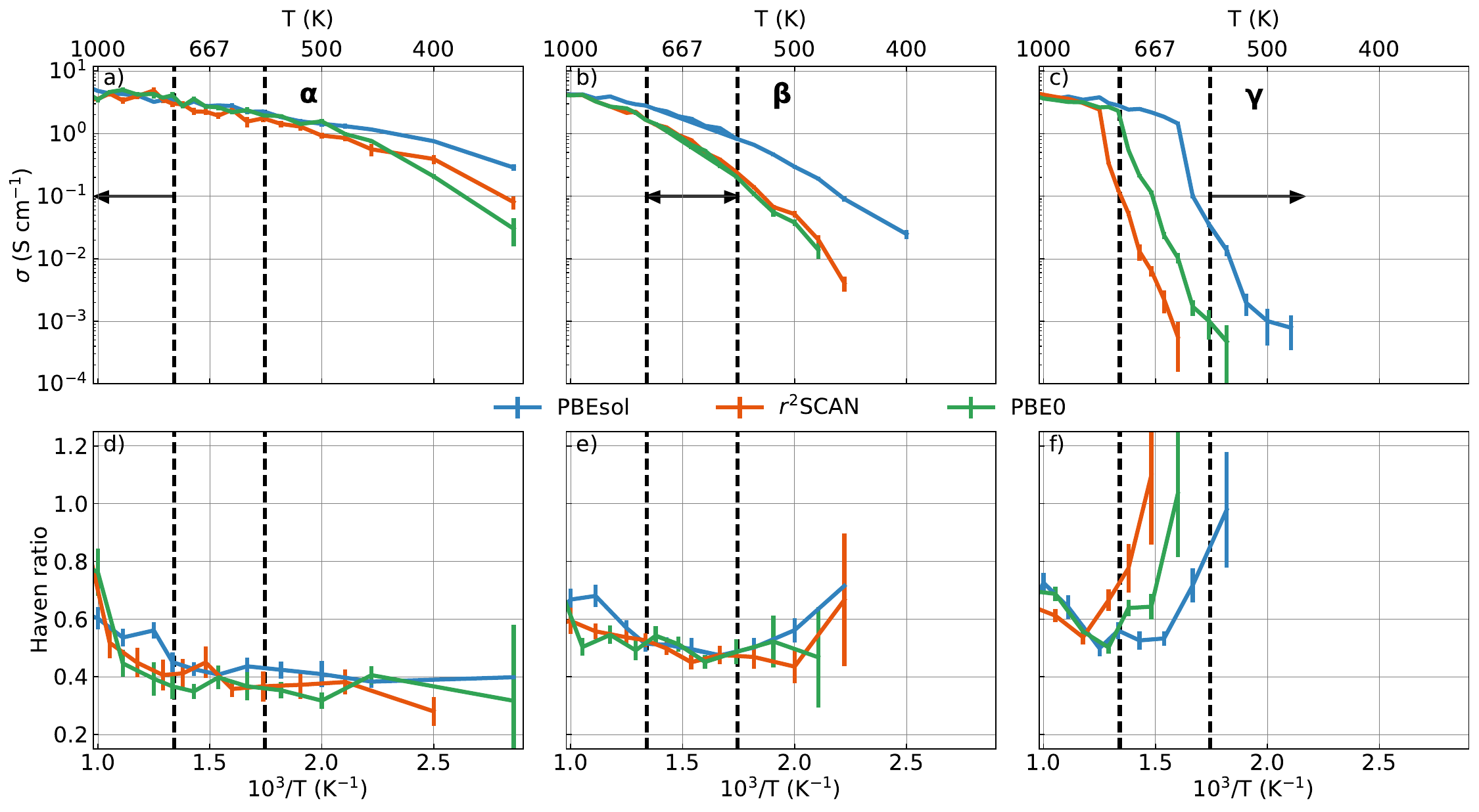}
    \caption{Temperature dependence of $\sigma$ and Haven Ratio. {Panels a), b) and c)} show a comparison between the {ionic} conductivities predicted by the ML models as a function of the inverse temperature, computed via the Green-Kubo relation. {Panels d), e) and f)} show the behaviour of the Haven ratio, $H_R = \sigma_{\textrm{NE}}/\sigma_{\textrm{GK}}$, as a function of the inverse temperature. Error bars obtained from standard block analysis over eight blocks are displayed. Vertical dashed lines indicate the experimental stability boundaries for the three phases. }
    \label{fig:Haven-ratio}
\end{figure*}

The upper panels of Fig.~\ref{fig:Haven-ratio} shows the temperature dependence of the {ionic} conductivity at zero pressure for a set of NpT simulations that start from the ideal $\alpha$, $\beta$ and $\gamma$ polymorph. 
Imperfect ergodicity, and the constraints on cell shape, make simulations dependent on the initial conditions. Even though simulations can only be considered converged within the stability range of each phase and target functional, results outside this range still report useful information about their behavior when metastable.
Thus, results are shown for every target functional and for all temperatures where the {ionic} conductivity is non-zero within the errorbars.

All the ML models predict the $\gamma$/$\alpha$ phase to be the least/most conductive, while the $\beta$ phase has intermediate values of $\sigma$ over a wide temperature window (up to 600-800 K depending on the model). This result is in agreement with previous computational studies \cite{yang_first-principles_2015, Staacke2022} and experimental measurements of the {ionic} conductivity \cite{HOMMA201153, kaup_impact_2020} on the known crystalline phases of \LiPS.  The negative slope of the profiles of $\sigma$ with respect to the inverse temperature is typical of the Arrhenius plots. In fact, since lithium-ion diffusion occurs via thermal activation, one can mathematically relates the Nernst-Einstein conductivity $\sigma_{\textrm{NE}}$ with the activation energy of the Li-hopping process. In particular, higher negative slopes correspond to higher activation energies. The $\gamma$ ($\alpha$) phase is thus not only the least (most) conductive phase but also the one that has the largest (smallest) activation barrier for lithium diffusion.

The high-temperature ends of the conductivity profiles of Fig.~\ref{fig:Haven-ratio} (for $T<670\,\textrm{K}$ for ML-PBEsol, $T < 880\,\textrm{K}$ for ML-$\textrm{r}^2$SCAN and  $T < 770\,\textrm{K}$ for ML-PBE0) give us additional insights. While the curves related to the three phases clearly span different conductivity ranges and show different slopes at low temperatures, this difference is not noticeable any more at high temperatures and the conductivity and the activation energies all approach the values of the $\alpha$ phase, with a characteristic kink that is most visible for simulations initialized in the $\gamma$ phase. The critical temperature at which this kink occurs depends on the reference functional. This is a clear effect of the structural transition studied in Sec. \ref{sec:phase-trans} and further investigated in Sec. \ref{sec:PS4-rotations} by an analysis of the \ce{PS4} rotational dynamics. In other words, the \ce{PS4} flips that induce the transition from the $\gamma$ polymorph to the partly-ordered $\beta$-$\alpha$ structure are responsible for the changes of the ionic conductivity due to the larger availability of hopping sites for Li-ions \cite{HOMMA201153}, as well as a reduction of the Li-hopping activation barrier. This conclusion is also highlighted by simulations of the $\beta$ and $\alpha$ phases at low temperatures, that show large $\sigma$ although \ce{PS4} flips are suppressed and librations are weak.
To quantify the reduction of the activation energy due to the structural transition, in \cref{tab:activation-en-gamma}, we report the activation energies that are fit to the Nernst-Einstein conductivities below and above the transition temperatures of each model (see {Sec.~X} of the Supporting Information for additional details). 
The effect of the transition is remarkable, as we observe a reduction of the activation energies of up to a factor of 6 depending on the reference DFT functional. Furthermore, the activation energies are very small above the transition, with values ranging between 0.25 and 0.32 eV. These values are very close to the values observed for the $\alpha$ phase and smaller than the $\beta$ phase (see \cref{tab:activation-en}), indicating a superionic behavior. In contrast, we note that the transition to the molten salt, that we observed for $T > 800$\,K in Sec.~\ref{sec:PS4-rotations} has practically no effect on the conductivity profiles.

This analysis, combined with the results of Secs.~\ref{sec:phase-trans} and \ref{sec:PS4-rotations}, also allows us to rule out any paddle-wheel effect, whereby \ce{PS4} motion is time-correlated with Li-hopping and increases Li-ion diffusion. In fact, due to the different rates of \ce{PS4} flipping ($\approx$ one every ns) and Li-ion hopping ($\approx$ one every ps) even at large temperature right below melting, the two mechanisms cannot be related. This point is strengthened in {Fig.~S12 of the Supporting Information,} showing the fast and linear increase of the mean square displacement of lithium ions in the simulation at 750 K of Fig. \ref{fig:Euler-dyn} over a 30-ps time window. 
{Instead, the interaction between Li-ion diffusion and \ce{PS4} libration, which share the same time scale, may be important, as we have directly inspected in Sec.~\ref{sec:spatial-corr} by analyizing the local contributions to $\sigma$ stemming from the interaction between Li and S ions.}

The lower panels of Fig.~\ref{fig:Haven-ratio} show the temperature dependence of the Haven ratio, $H_R$, computed using Eq.~\eqref{eq:HR}. As anticipated in Sec. \ref{sec:GKtheory}, this coefficient quantifies the discrepancy between $\sigma$ and $\sigma_\mathrm{NE}$. 
$H_R<1$ for almost every temperature and only for the $\gamma$ phase it approaches one at low temperatures, where the system is weakly conductive. This indicates the presence of inter-atomic correlations in the ionic conductivity, both between carriers (Li$^+$) and the solid matrix (\ce{PS4}), that cannot be captured by the NE approach (see Eq.~\eqref{eq:NE}), as the latter only estimates the conductivity based on the self-diffusion of the lithium ions. This effect is most pronounced in the $\alpha$ phase, where $H_R \approx 0.4$ below melting, meaning that the NE estimate underestimates the GK conductivity by more than a factor of two. At high temperatures, the Haven ratio slightly increases, indicating that the material becomes disordered. This might be a result of the phase transformations of the solid matrix, which we expect to weaken the interionic correlations. Still, the Haven ratio never exceeds 0.8, even at $1000~$K for any of the ML models studied.

\begin{table}
    \centering
    \begin{tabular}{cccc}
        \toprule 
        model & $E_A(T<T_c)$ (eV)  & $E_A(T>T_c)$ (eV) & ratio \\
        \midrule 
        ML-PBEsol & $0.93 \pm 0.07$
        & $0.249 \pm 0.004$ & $3.4 \pm 0.3$ \\
        ML-$\text{r}^2$SCAN & $1.64 \pm 0.05$ & $0.32 \pm 0.01$ & $5.1 \pm 0.2$ \\
        ML-PBE0 & $1.43 \pm 0.06$ & $0.269 \pm 0.007$ & $5.3 \pm 0.3$ \\
       \bottomrule 
    \end{tabular}
    \caption{Activation energies for Li-ion diffusion for the simulations initialized in the $\gamma$ phase for temperatures below the phase transition temperature ($T > T_c$) observed in Sec.~\ref{sec:PS4-rotations} and above $T_c$, see the SI for details on the computation of the activation energies. The last column is the ratio between the $E_A(T<T_c)$ and $E_A(T>T_c)$ and quantifies the reduction of the Li-diffusion activation energy due to the phase transition.}
    \label{tab:activation-en-gamma}
\end{table}

\begin{table*}
    \centering
    \begin{tabular}{cccccc}
          \toprule 
         & $E_A^{\beta}$ (eV) & $E_A^{\alpha}$ (eV) & $\sigma^{\beta}_{298K}$ (S $\textrm{cm}^{-1}$) & $\sigma^{\beta}_{500K}$ (S $\textrm{cm}^{-1}$)  & $\sigma^{\alpha}_{298K}$ (S $\textrm{cm}^{-1}$)\\
       \midrule
        PBEsol & 0.38 & 0.17 & 7.7$\times 10^{-4}$ & 2.8$\times 10^{-1}$ & 1.9$\times 10^{-1}$ \\
        $\text{r}^2$SCAN & 0.57 & 0.21 & 2.1$\times 10^{-5}$ & 
        5.1$\times 10^{-2}$& 9.2$\times 10^{-2}$ \\
        PBE0 & 0.62 & 0.19 & 8.7$\times 10^{-6}$& 3.9$\times 10^{-2}$& 1.6$\times 10^{-1}$ \\
        exp. & 0.47 \cite{liu_anomalous_2013}, 0.36 \cite{kaup_impact_2020} & 0.22 \cite{kaup_impact_2020} & 8.9$\times 10^{-7}$ \cite{liu_anomalous_2013,self_solvent-mediated_2020} & 3.0$\times 10^{-2}$ \cite{HOMMA201153}&  \\
        AIMD-PBE & 0.40 \cite{de2018analysis} & 0.18 \cite{kim2018thermally} & & & 8.0$\times 10^{-2}$\cite{kim2018thermally} \\
       \bottomrule
    \end{tabular}
    \caption{Comparison between the predicted activation energies and conductivities (computed at 500 K and extrapolated at room temperature; see SI for details) of the $\beta$ and $\alpha$ phases with both experimental references \cite{liu_anomalous_2013,kaup_impact_2020} and previous \textit{ab initio} MD studies \cite{de2018analysis,kim2018thermally,yang_elastic_2016}. }
    \label{tab:activation-en}
\end{table*}

In \cref{tab:activation-en} we quantitatively compare the activation energies and the ionic conductivities predicted by our ML models with recent experimental measurements \cite{liu_anomalous_2013,kaup_impact_2020} and with computational studies based on AIMD at the PBE level \cite{de2018analysis,kim2018thermally,yang_elastic_2016}. The first two columns of \cref{tab:activation-en} compare the activation energies in the $\beta$ and $\alpha$ phase, while the remaining three report the estimates of the ionic conductivity of the $\beta$ phase at 298 K\footnote{This value is extrapolated from a fit of the temperature profiles of the $\beta$ structure (see Fig. \ref{fig:Haven-ratio}) for each ML model, see SI for details on the computation of the activation energies.} and 500 K and the ionic conductivity of the $\alpha$ phase at 298 K. The ML-PBEsol model predicts activation energies in agreement with previous AIMD for both the $\beta$ and the $\alpha$ structure. In the case of the $\beta$ phase, the ML-$r^2$SCAN and ML-PBE0 predict an activation energy that is greater than the one computed with the ML-PBEsol model, but overall the values are close to the experimental results. For the $\alpha$ phase the activation energies are particularly close to the experiment.
These last results are remarkably good in particular when comparing them with the prediction of the classical empirical potential recently introduced by Forrester et \textit{al}~\cite{forrester2022}. For this model, the activation energy for stoichiometric $\alpha$-\LiPS~is much larger ($0.40$ eV) than the experimental result and similar to the value of the $\beta$ phase. The empirical potential also predicts the $\alpha$ phase to be only slightly more conductive than the $\beta$ phase for all temperatures. 

In conclusion, our analysis shows that the ML potentials are the only possible solution to accurately predict the properties of \LiPS, given the unreliability of empirical potentials and the prohibitive cost of \textit{ab initio} simulations, in particular at PBE0 level.

\subsection{Spatial correlations}
\label{sec:spatial-corr}
In this last section, we investigate the role of local correlations in determining the full {ionic} conductivity by computing the spatial dependence of the integral of the partial cross-correlation functions, $I_{\mathrm{Li}A}$, between lithium atoms and other atomic species $A=\mathrm{Li,P,S}$. 
However, before we start this analysis, it is necessary to make a few technical considerations. While the total conductivity does not depend on the frame of reference, due to the charge neutrality of the simulation cell, the value of any partial correlation does depend on it, as we show in the SI. For instance, in the reference frame of the matrix, all the Green-Kubo integrals of the partial correlation between Li and the other species (P and S) vanish.
In contrast, in the reference frame of the center of mass of the entire system the solid matrix recoils due to the diffusion of the center of mass of Li atoms (see SI).
To carry out our analysis, we choose the reference frame where the center of mass of the PS solid matrix remains stationary, e.g. $\sum_{i\in \mathrm{P, S}}\mathbf{v}_i=0$ at every instant. This choice is motivated by the fact that this reference frame is in principle the same in which the lithium diffusivity, entering the NE relation, Eq.~(\ref{eq:NE}), should be computed. Moreover, this is the most natural choice for a battery, since in this reference frame the solid matrix of the battery is not moving.

To study the spatial dependence of the integral of the correlation functions we perform a Gaussian kernel density estimation (KDE) \cite{ked} of the correlation functions: 
\begin{align}
    I_{\mathrm{Li}A}(r)  \equiv \, &  \frac{\sum_{i \in \mathrm{Li}}\sum_{j \in A}\int_{0}^{\infty}  C_{ij}(t,r)\, dt}{\sum_{i \in \mathrm{Li}}\sum_{j \in A}{\bar{w}_{ij}(r)}}  \label{eq:ICCF} \\
    C_{ij}(t,r) \equiv \, & \tfrac{1}{3} \left\langle \mathbf{v}_{i}(t) \cdot \mathbf{v}_j(0)\, w_{ij}(0,r) \right\rangle \label{eq:CCFij} \\
    w_{ij}(t,r) \equiv \, & \frac{1}{\sqrt{2\pi \varsigma^2}} \exp\left[-\tfrac{\left(r_{ij}(t)-r\right)^2}{2\varsigma^2}\right]\label{eq:wij} 
\end{align}
where $i$ runs over all Li atoms and $j$ runs over all the atoms of type $A$; $w_{ij}(t,r)$ is a Gaussian weight with width $\varsigma$, which we set to 0.33\AA; $\bar{w}_{ij}(r)$ is the time average of $w_{ij}(r)$ over the whole trajectory. 
The connection between \cref{eq:ICCF} and the {ionic} conductivity of Eq.~\eqref{eq:GKeq} is readily established. In fact, within the KDE,
\begin{equation}
     \frac{\Omega}{N_{\mathrm{Li}} N_A}\sum_{i \in \mathrm{Li}}\sum_{j \in A}\bar{ w}_{ij}(r) \underset{\varsigma \to 0}{\approx} 4 \pi  r^2 \,g_{\mathrm{Li}A}(r),
\end{equation}
where $g_{\mathrm{Li}A}(r)$ is the radial distribution function between the $\mathrm{Li}$ and the species $A$. 
{Therefore, the integral over $r$ of $I_{\mathrm{Li}\mathrm{Li}}(r)$ plays the role of the correction to the NE relation that is needed to recover the full GK conductivity:
\begin{align}
   \sigma -\sigma_\mathrm{NE}&\propto  \int_{0}^{\infty} I_{\mathrm{Li}\mathrm{Li}}(r)  g_{\mathrm{Li}\mathrm{Li}}(r) 4 \pi r^2 dr 
\end{align}
\cref{fig:spatial correlation} shows $I_{\mathrm{Li}A}(r)$ for different values of $r$. The most relevant correlations come from the Li-Li and Li-S interactions, while the correlations with the P are compatible with zero for any $r$. As expected, for large $r$ all the correlations become compatible with zero. The oscillations in the correlations functions follow, as expected, the peaks in the radial distribution function $g_{\mathrm{Li}A}(r)$ (lower panel), and show that only the first shell of Li atoms and the first two of S atoms are important for the conductivity. 
The correlation of the first shell of Li is strongly positive in agreement with Ref.~\cite{he2017origin}, where it was shown that the hopping of Li atom in one direction facilitates the movement of the other Li in the same direction.

\begin{figure}
    \centering
    \includegraphics[width=\columnwidth]{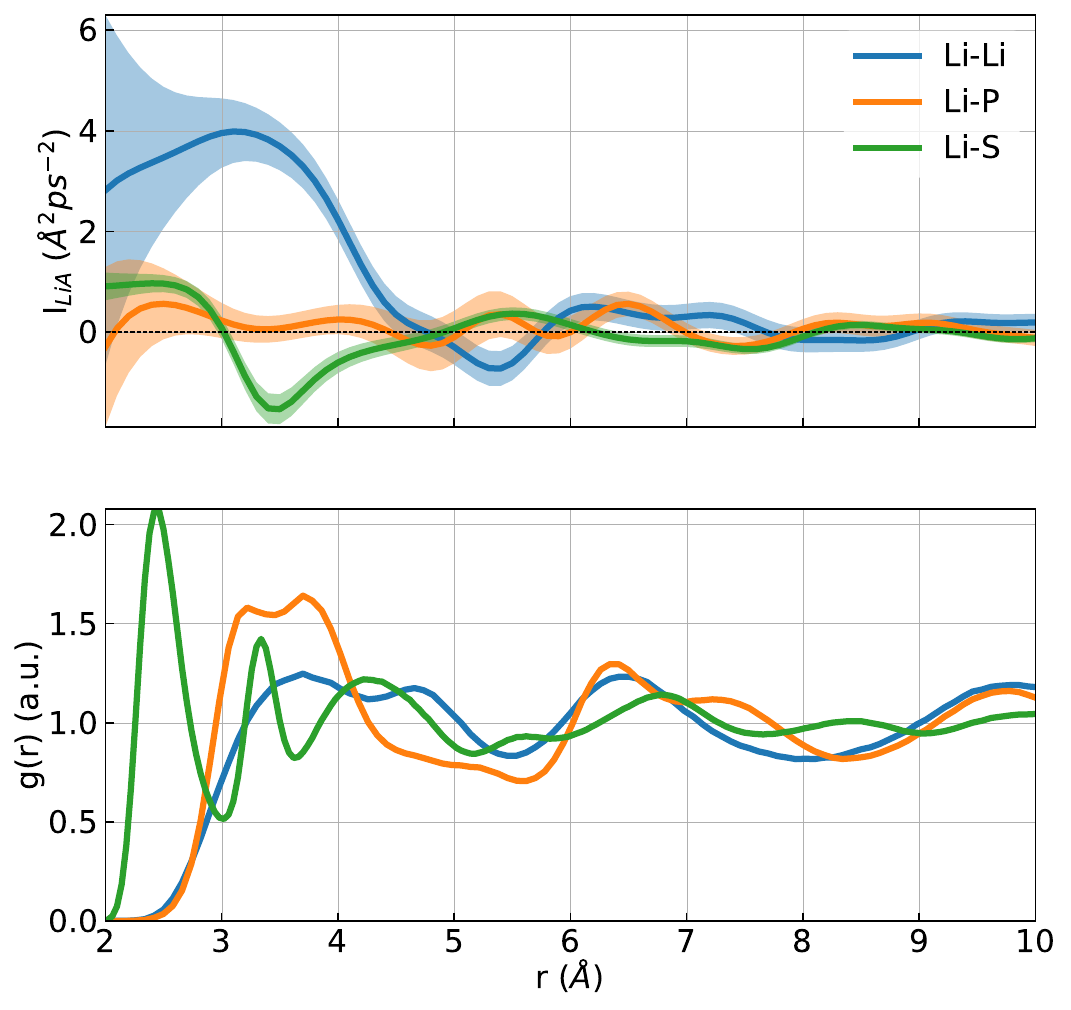}
    \caption{(Upper panel) Integrals of the cross correlation functions as defined in \cref{eq:ICCF}, for different pairs of atomic species and as a function of $R$. All the velocities are computed in the system of reference of the center of mass of the solid matrix of the battery. The shaded area indicate the uncertainty on the mean value, obtained from block analysis on 10 blocks.
    (Lower panel) Radial distribution function between the Li and all the other atomic species. The data reported in the figure are obtained from 190 ps of a simulation of the $\alpha$-phase at 650 K.}
    \label{fig:spatial correlation}
\end{figure}

\section{Conclusions}

In this work, we have presented a computational study of lithium ortho-thiophosphate via multiple machine learning models, targeting DFT references of increasing accuracy, and elucidating the critical role of \ce{PS4} flips and phase transitions in determining the observed superionic behavior of \LiPS. We find that all the ML models predict two distinct phase transitions. First, we observe a transition from the $\gamma$ to a partly ordered phase with both $\beta$ and $\alpha$ alignments, that cannot be fully resolved at the time and length scale of the MD simulations. While this limitation does not allow to provide a prediction of the full phase diagram of \LiPS, we identify the presence of collective \ce{PS4} flips as the driving mechanism. Secondly, we observe the melting of the system into its constituent $\textrm{Li}^{+}$ cations and $\textrm{PS}_4^{3-}$ polyanions at elevated temperatures ($>$800 K).  Both these transitions are associated with drastic changes in the rotational dynamics of the \ce{PS4} groups as a function of the temperature and appear as distinct peaks of the heat capacity.

We also compute the {ionic} conductivity of \lips~in all its stable polymorphs and elucidate the importance of including the effects of interatomic correlations, by computing it with the full Green-Kubo theory of linear response and with the Nernst-Einstein approximation. We find that the interionic correlations account for considerable deviations between the NE and the GK estimates, as quantified by a Haven ratio that is smaller than one in every polymorph at all temperatures, except for the weakly conductive $\gamma$ structure at low temperatures. Notably, the Haven ratio reaches values of 0.4 in the highly conductive $\alpha$-phase, suggesting that a pure NE approach can result in the underestimation of the conductivity by more than a factor of two. From a spatially-resolved analysis performed in the reference frame of the solid matrix, we find that most of these interionic correlations come from the first shell of Li-Li neighbors, thus indicating that a concerted Li-ion hopping is a key aspect of charge transport in this material, in agreement with Ref.~\cite{he2017origin}. 

Finally, we investigate how the observed phase transitions of \LiPS~affect the ionic conductivity. We find that the occurrence of correlated \ce{PS4} flips results in a dramatic decrease of the activation energy (up to a factor of six) when the system transitions from the $\gamma$ to a mixed $\beta$-$\alpha$ phase. Furthermore, we show that subsequent \ce{PS4} flips occur at the time scale of nanoseconds, that is much larger than the typical time laps between two subsequent lithium ion hoppings, in agreement with Refs.~\cite{he2017origin}. We thus conclude that the sudden change in the PES of the lithium ions that is due to the rearrangement of the \ce{PS4} tetrahedra is the physical mechanism for the observed superionic behavior of \LiPS. Crucially, this mechanism is fundamentally different from the one proposed in Ref. \cite{zhang_targeting_2020}. There, a characteristic paddlewheel effect was observed in AIMD-PBE simulations at elevated temperatures and invoked to explain the neutron diffraction measurements showing a polyanion reorientational disorder. The AIMD simulations were however limited in size and thus showed much larger finite-size effects than the ones we observe in this work, including an artificial stabilization of the solid phase up to temperatures (1200 K) that are much larger than the nominal melting point of \LiPS. We also stress that the paddlewheel effect itself was observed in AIMD simulations at these very high temperatures, thus likely making it an artifact of the small simulation box that was used there.
The ML models that we present in this work overcome these limitations and offer a more natural interpretation of the experimental results. This finding also suggests additional directions of research in the quest for a promising solid electrolyte and potentially a way to design new target compounds.  In particular, we expect that tentative modifications of \LiPS~to stabilize its superionic phases at room temperature, by e.g. atomic substitution and amorphization, should be accompanied by a reduction of the polyanion rotational free energy barrier, that limits the spatial extension of the \ce{PS4} fluctuations. Further developments of this work thus imply a detailed thermodynamic study of the phase transitions observed here and a comparison of multiple different SE compounds, with the aim of suggesting a target compound for experimental synthesis.

While providing these useful mechanistic insights, our ML models show a remarkable agreement with experiment in the prediction of a number of independent quantities. 
Specifically, the PBE0 functional provides the best agreement on the prediction of the electronic band gap, while the associated ML-PBE0 model reaches overall the best accuracy on the prediction of lithium activation energies in the $\beta$ and $\alpha$ phase, the {ionic} conductivities at 298 and 500\,K and the melting temperature. In particular, our results proved to be much more accurate than empirical potentials \cite{forrester2022}, that are often used to overcome the high computational cost of AIMD. 
These results and the observed dependence of the finite-temperature predictions of the ML models on the DFT reference indicate the necessity of using more accurate functionals for the description of transport properties in solid electrolytes, than state-of-the-art GGA functionals. Machine learning becomes then a necessary step in modelling this class of materials, as \textit{ab-initio} studies with the PBE0 functional are far beyond reach because of their very high computational cost.

In conclusion, we have shown how the use of machine learning potentials for a prototypical solid electrolyte captures the mechanisms of the transition to its superionic phase and the quantitative values of the ionic conductivity, while also allowing us to investigate the role of inter-ionic correlations.
This work thus opens up a new frontier in the exploration of superionic materials as it allows their large-scale simulations at hybrid-DFT accuracy for hundreds of nanoseconds. This will offer crucial insights into the fundamental properties of solid electrolytes, as well as guidance for the experimental realization of new candidate compounds.

\section*{Supporting information}

{The Supporting Information is organized as follows:
Sec.~I contains the parameters of the ML models and the learning curves \cite{Bartok2010,Deringer2021,Zaspel2019,librascal,skmatter}.
Sec.~II shows the results from kernel principal component analysis \cite{scholkopf1997kernel,PhysRevB.105.165141}.
Sec.~III contains the analysis of the local prediction rigidity \cite{chong2023robustness}.
Sec.~IV shows the comparison between the mean square displacement from an \textit{ab initio} simulation and a simulation with the ML-PBEsol model.
Sec.~V contains the details of the DFT calculations \cite{gian+09jpcm, giannozzi_advanced_2017,Urru2020,CarmineoQE,PerdewPBEsol,VASP1,VASP2,VASP3,BlochPAW,Kresse1999,car_unified_1985,nose_unified_1984,RRKJ}.
Sec.~VI shows the band structure of the $\gamma$ phase.
Sec.~VII contains the analysis of a heat-quench trajectory.
Sec.~VIII shows the behavior of the order parameter during a collective plane flipping.
Sec.~IX contains the study of the $g(r)$ during melting.
Sec.~X shows results for the Nernst-Einstein approximation to the ionic conductivity.
Sec.~XI contains theoretical insights \cite{Marcolongo2017,Marcolongo2016,Grasselli2021} on the frame dependence of the breakdown of the electrical conductivity into species-dependent terms.}
\begin{acknowledgments}

We thank M.~L.~Kellner for a critical review of an early version of this manuscript and S.~Lombardi for technical help in improving the figures. L.G., D.T. and M.C. acknowledge funding from the Swiss National Science Foundation (SNSF) under the Sinergia project CRSII5\_202296 and support from the MARVEL National Centre of Competence in Research (NCCR) for computational resources. F.G. and M.C. acknowledge funding from the European Research Council (ERC) under the European Union’s Horizon 2020 research and innovation programme Grant No.~101001890-FIAMMA. F.G.~also acknowledges funding from the European Union's Horizon 2020 research and innovation programme under the Marie Sk\l{}odowska-Curie Action IF-EF-ST, grant agreement No.~101018557-TRANQUIL. This work was supported by grants from the Swiss National Supercomputing Centre (CSCS) under the projects s1092 and s1219 and by the Platform for Advanced Scientific Computing (PASC) project ``Machine learning for materials and molecules: toward the exascale''.

\end{acknowledgments}

\section*{Declaration of interests}
The authors declare no competing interests.

\section*{Data availability statement}
The \LiPS~datasets and the ML-potentials generated in this work, as well as the input files of the simulations, the notebooks and the scripts employed are available on the Materials Cloud Platform \cite{talirz2020materials} at \href{https://doi.org/10.24435/materialscloud:g2-fp}{https://doi.org/10.24435/materialscloud:g2-fp}.
\bigbreak

\def\bibsection{}  

\centerline{ \textbf{\small{REFERENCES}} }
\bigbreak
\end{document}